\documentclass[sigconf, preprint]{acmart}

\AtBeginDocument{%
  }

\usepackage{booktabs} 
\usepackage{tabularx} 
\usepackage[normalem]{ulem} 
\useunder{\uline}{\ul}{}

\begin{document}

\title{HypomimiaCoach: An AU-based Digital Therapy System for Hypomimia Detection \& Rehabilitation with Parkinson's Disease}

\author{Yingjing Xu}
\authornotemark[1]
\email{poppyxu@zju.edu.cn}
\affiliation{
  \institution{Zhejiang University}
  \country{China}
}

\author{Xueyan Cai}
\authornotemark[1]
\email{22251378@zju.edu.cn}
\affiliation{
  \institution{Zhejiang University}
  \country{China}
}

\author{Zihong Zhou}
\email{zihongzhou@zju.edu.cn}
\affiliation{
  \institution{Zhejiang University}
  \country{China}
}

\author{Mengru Xue}
\email{mengruxue@zju.edu.cn}
\affiliation{
  \institution{Zhejiang University}
  \country{China}
}

\author{Bo Wang}
\email{wangke1121@163.com}
\affiliation{
  \institution{Department of Neurology, the Second Affiliated Hospital}
  \country{China}
}

\author{Haotian Wang}
\email{wanght816@zju.edu.cn}
\affiliation{
  \institution{Department of Neurology, the Second Affiliated Hospital}
  \country{China}
}

\author{Zhengke Li}
\email{zhengk.li@qq.com}
\affiliation{
  \institution{Zhejiang University}
  \country{China}
}

\author{Chentian Weng}
\email{22351214@zju.edu.cn}
\affiliation{
  \institution{Zhejiang University}
  \country{China}
}

\author{Wei Luo}
\email{luoweirock@zju.edu.cn}
\affiliation{
  \institution{Department of Neurology, the Second Affiliated Hospital}
  \country{China}
}

\author{Cheng Yao}
\email{yaoch@zju.edu.cn}
\affiliation{
  \institution{Zhejiang University}
  \country{China}
}

\author{Bo Lin}
\authornotemark[2]
\email{rainbowlin@zju.edu.cn}
\affiliation{
  \institution{Zhejiang University}
\institution{Binjiang Institute of Zhejiang University}
  \country{China}
}

\author{Jianwei Yin}
\email{zjuyjw@cs.zju.edu.cn}
\affiliation{
  \institution{Zhejiang University}
  \country{China}
}

\renewcommand{\shortauthors}{Xu et al.}

\begin{abstract}
Hypomimia is a non-motor symptom of Parkinson's disease that manifests as delayed facial movements and expressions, along with challenges in articulation and emotion. Currently, subjective evaluation by neurologists is the primary method for hypomimia detection, and conventional rehabilitation approaches heavily rely on  verbal prompts from rehabilitation physicians. There remains a deficiency in accessible, user-friendly and scientifically rigorous assistive tools for hypomimia treatments. To investigate this, we developed HypomimaCoach, an Action Unit (AU)-based digital therapy system for hypomimia detection and rehabilitation in Parkinson's disease. The HypomimaCoach system was designed to facilitate engagement through the incorporation of both relaxed and controlled rehabilitation exercises, while also stimulating initiative through the integration of digital therapies that incorporated traditional face training methods. We extract action unit(AU) features and their relationship for hypomimia detection. In order to facilitate rehabilitation, a series of training programmes have been devised based on the Action Units (AUs) and patients are provided with real-time feedback through an additional AU recognition model, which guides them through their training routines. A pilot study was conducted with seven participants in China, all of whom exhibited symptoms of Parkinson's disease hypomimia. The results of the pilot study demonstrated a positive impact on participants' self-efficacy, with favourable feedback received. Furthermore, physician evaluations validated the system's applicability in a therapeutic setting for patients with Parkinson's disease, as well as its potential value in clinical applications.

\end{abstract}






\maketitle

\section{Introduction}
Parkinson's disease is a chronic neurological disorder that requires long-term treatment and management, primarily affecting people aged 60 years and older \cite{kalia2015parkinson}. It is estimated that by 2030, the number of PD patients in China will increase to form over half of the world's PD patients \cite{dorsey2007projected}. Research into more accessible and patient-friendly treatments is needed to reduce the burden on healthcare systems and individual patients. Hypomimia is a non-motor symptom of Parkinson's disease characterized by deficits in facial expression, mainly in the form of impaired facial expression or reduced facial movements in patients \cite{maycas2021hypomimia}. It can also cause a range of complications, including language deficits (dysarthria, disruption of fluency, etc.), which are often accompanied by symptoms of depression, mood swings, and cognitive impairment.

Existing researchers have devised various metrics to help physicians assess the severity of Parkinson's disease hypomimia \cite{pegolo2023comparison, pegolo2022quantitative, grammatikopoulou2019detecting}, and a widely used clinical diagnostic method is the Movement Disorder Society-Unified Parkinson's Disease Rating Scale (MDS-UPDRS) published by the World Federation of Neurology \cite{goetz2008movement}. However, traditional clinical judgment of hypomimia symptoms in Parkinson's disease relies heavily on in-person assessment by an experienced neurologist \cite{bianchini2024story}. This approach often faces time and space constraints that may prevent timely disease detection and intervention. In addition, limited healthcare resources hinder the ability of patients to effectively self-assess and rehabilitate their hypomimia. The use of artificial intelligence (AI)-based algorithms to detect hypomimia provides a convenient way to diagnose Parkinson's disease, thereby actually facilitating early detection and intervention of hypomimia symptoms \cite{lee2021human, wootton2019unmoving}. Previous studies have extensively explored the use of AI models in Action Unit(AU) \cite{ekman1978facial} recognition \cite{liu2020relation, li2019semantic}, achieving high accuracy in labeling facial action units. Despite this potential, there are substantial gaps in research on the clinical applications of these quantifiable AUs in diagnosis and rehabilitation. Therefore, it is crucial to develop more objective and accurate assessment tools for facial rehabilitation and to explore strategies for seamlessly integrating AI techniques into facial rehabilitation training.

Targeted training of facial and tongue muscles for Parkinson's patients can effectively alleviate facial stiffness. Traditional methods of facial training \cite{kang2022effectiveness, kang2022effectiveness, karp2019facial} require patients to voluntarily exercise their facial muscles, and patient autonomy and participation in the training is critical. However, in current clinical practice, adherence to these exercises is often challenging and patients vary in their condition and the severity of their facial symptoms. Consistent execution of the exercises cannot be guaranteed. Digital therapies offer Parkinson's patients more flexible, convenient, and accessible rehabilitation options for transitioning from the clinical setting to everyday life than traditional rehabilitation methods, with the potential to improve recovery outcomes and overall quality of life. Digital therapies can psychologically alleviate tension, anxiety, and depression in patients, increasing acceptance and engagement. Researchers have explored digital treatments for facial disorders. For example, Liu et al. \cite{liu2018pumpkin} developed a Parkinson's hypomimia game, which was explored and experimented with as a treatment. However, there is still a lack of research on the clinical application of digital therapeutic tools for hypomimia that can be widely used, and people with Parkinson's disease may face problems with the accessibility of specialized and systematic digital therapeutic tools, especially in resource-limited areas. 

Self-determination theory \cite{deci2012self} states that the motivation to use a product depends on the extent to which the product fulfills the basic needs of autonomy, relevance, and competence. Systems capable of artificial intelligence combined with traditional medical rehabilitation therapy can increase relevance and also give the user more autonomy. Thus, allowing users to interact through familiar, easy, and novel interfaces may be a solution to increase user engagement and treatment outcomes. 
Self-efficacy is an important belief that needs to be improved in people with Parkinson's disease because it enables individuals to develop confidence in their ability to control their motivation, behavior, and social environment. High self-efficacy prevents the negative effects of emotionally oriented coping. Improving self-efficacy in the early stages of Parkinson's disease after masked hypomimia may have beneficial long-term effects. Digital interventions can positively impact self-efficacy in people with Parkinson's by promoting positive behavioral change and increasing confidence in managing symptoms. Digital technologies, such as mobile apps and online platforms, are developed and utilized to provide patients with easy access to rehabilitation resources and support. These technologies can provide timely feedback, progress tracking, and interactive training, and this flexibility and personalized attention can further enhance patients' self-efficacy. Therefore, this study aims to explore the use of digital interventions in the rehabilitation of patients with PD hypomimia and how they can be a potentially valuable addition to existing research on digital therapies.

Overall, we sought to answer the following research questions.

\begin{itemize}
\item \textbf{RQ1}: How can artificial intelligence techniques be applied to help patients diagnose and treat Parkinson's hypomimia symptoms?

\item \textbf{RQ2}: How to design interactive tools for digital therapies applicable to patients with Parkinson's hypomimia symptoms?
\end{itemize}

To answer these questions, the researchers in this thesis designed a detection and rehabilitation training system based on facial action expression units (AUs), relying on existing artificial intelligence facial action unit recognition technology. The detection process includes four steps: data acquisition and preprocessing, facial AU feature extraction, facial AU feature fusion, and hypomimia classification. Parkinson's disease hypomimia faces characterized by reduced facial movements and expressions can be captured more accurately by Action Units (AU) analysis. Graph Convolutional Networks (GCNs) further improve the model performance by utilizing the interconnectivity of AUs. Based on a comprehensive review of existing facial rehabilitation techniques, we present a novel interactive training system utilizing action units (AUs). By combining targeted exercises for the brow, eye muscles, cheeks, nose, lips, and articulation with AI-based feedback, we develop a digital therapy with real-time feedback that enables users to actively participate in their rehabilitation process. The method quantifies the extent of facial movements, which allows for better assessment of facial movements in Parkinson's patients and helps physicians better develop rehabilitation programs. Combined with traditional facial training methods, this system aims to improve patient autonomy, relevance, and competence, thus potentially contributing to the patient's rehabilitation outcome.

In summary, this paper has three main contributions:

\begin{itemize}
\item We extend traditional detection and rehabilitation techniques by combining ai technology (i.e., AU facial recognition analysis models) with traditional medical treatments through multisensory interaction and quantitative real-time feedback.

\item HypomimiaCoach, an AU-based digital therapy system, for low-imagery detection and rehabilitation of Parkinson's disease patients. The detection component employs a GNN model to capture AU features and their interrelationships to facilitate the recognition of hypomimia. The rehabilitation component combines non-volatile recognition with facial rehabilitation training, utilizing a non-volatile recognition model to help users with basic and advanced training. 

\item We conducted a user study with 7 patients and 10 physicians at a hospital in China to evaluate the usability of the system, focusing on participant engagement, satisfaction, and initiative.
\end{itemize}
\section{Related Work}
\subsection{Clinical Facial Detection and Rehabilitation for Parkinson's Disease}
A variety of rating scales have been designed to evaluate the severity of Parkinson's disease, including the Hoehn and Yahr staging scale \cite{hoehn1967parkinsonism}, the UK Queen Square Brain Bank Criteria \cite{gibb1988relevance}, the Montreal Cognitive Assessment \cite{nasreddine2005montreal}, and the MDS-UPDRS\cite{nasreddine2005montreal}. Nevertheless, the application of these scales is often constrained by the neurologist's experience and subjective assessment, as well as factors such as time and distance.

Clinical facial rehabilitation is a therapeutic intervention aimed at improving facial movement and expression control in individuals with facial muscle dysfunction. Thai massage, Japanese massage, Traditional Chinese Tuina, and some other manual therapies have clinical records demonstrating their ability to improve motor function in PD patients \cite{kang2022effectiveness}. Chen et al. \cite{kang2022effectiveness} introduced a passive massage rehabilitation exercise for hypomimia patients, reporting positive outcomes in facial expression and sleep quality. Karp et al.\cite{karp2019facial} demonstrated a strong association between facial rehabilitation and improved Facial Grading Scale (FGS) scores in facial nerve paralysis patients. Zheng et al.\cite{0Therapeutic} compared facial rehabilitation treatment with traditional medication and acupuncture for facial nerve neuritis, revealing enhanced effectiveness of facial paralysis rehabilitation exercises. This approach heavily relies on the physician's subjective evaluation, and traditional methods often lack real-time feedback and quantitative assessment mechanisms. Moreover, given limited healthcare resources, patients struggle to access scientific and professional guidance, hindering their ability to effectively self-assess and train their hypomimia.

\subsection{AI technologies for Hypomimia in Parkinson's Disease}
Artificial Intelligence (AI) systems are showing an increasing promise for numerous healthcare applications. Recently, the advantages of Deep Learning (DL) are spawning AI systems with human-like performance in several clinical domains \cite{calisto2023assertiveness, calisto2022breastscreening, hannun2019cardiologist, ruamviboonsuk2019deep, stephansen2018neural}. Artificial intelligence technologies have been applied in clinical treatment, medical research, and healthcare professional decision support. The utilization of artificial intelligence (AI)-based algorithms for hypomimia detection provides a convenient avenue for the diagnosis of Parkinson's disease, thereby enabling early detection and intervention. In previous research, Bandini et al.\cite{bandini2017analysis} extracted 49 facial keypoints and 20 facial features to detect hypomimia. Grammat et al. \cite{grammatikopoulou2019detecting} leveraged FaceAPI for keypoints extraction to construct linear regression model to detect. Su et al. \cite{su2021detection} utilized geometric and texture features extracted from facial images for detection. Later, Su et al. \cite{su2021hypomimia} trained an end-to-end model employing RGB image sequences and optical flow sequences in a dual-channel framework, incorporating temporal dynamic features. Recognizing the difficulties faced by Parkinson's patients in seeking in-person medical attention due to mobility issues and other factors \cite{su2021detection}, AI-driven Parkinson's detection facilitates remote diagnosis, alleviating temporal and spatial constraints in disease assessment.

The Facial Action Coding System (FACS) \cite{ekman1978facial} provides anatomically defined codes for precise characterization of facial expression changes, encapsulating various facial muscle movement states through Facial Action Units (AUs). AUs are the basic units of facial movement, each corresponding to a specific contraction or relaxation of a facial muscle or muscle group. By identifying and coding these AUs, FACS allows for a detailed and objective analysis of facial expressions, enabling researchers to study the relationship between facial expressions and various psychological states, emotions, and behaviors. Previous studies have extensively explored the utilization of AI models for AU recognition, achieving high levels of accuracy in labeling facial action units. Liu et al. \cite{liu2020relation} introduced a deep learning framework using Graph Convolutional Networks (GCN) for AU detection. Li et al. \cite{li2019semantic} explored semantic relationships among AUs in deep neural networks, enhancing feature representation in facial regions through the AU Semantic Relationship Embedding Representation Learning (SRERL) framework. Although AI models have advanced AU recognition capabilities, bridging the gap between these technological advancements and clinical rehabilitation practice remains a challenge. Action Units (AUs), which provide a quantitative measure of facial expressions, have seen significant advancements in their recognition through AI models. Despite this potential, there is a substantial gap in research regarding the clinical application of these quantifiable AUs in diagnosis and rehabilitation. Therefore, developing more objective and accurate facial rehabilitation assessment tools and exploring strategies for seamlessly integrating AI technologies into facial rehabilitation training is of paramount importance.

\subsection{Digital Therapy for Hypomimia in Parkinson's Disease}
  Digital therapies provide a more accessible treatment paradigm than traditional rehabilitation treatments and can psychologically alleviate patient tension, anxiety, and depression, increasing patient acceptance and engagement. Much of the current research in digital therapies has been used for early screening and rehabilitative interventions for diseases \cite{domHok2012break, wang2014eat, arshad2020east}. Researchers have explored digital therapies for facial disorders. Park st al. \cite{park2015facial} developed a facial expression training system using bilinear shape models, enabling users to match their expressions with 3D models, promising rehabilitation for various facial conditions. Barrios et al. \cite{barrios2021farapy} introduced FaraPy, a mobile augmented reality mirror therapy system for facial paralysis, preferred over traditional therapy. Liu et al. \cite{liu2018pumpkin} designed a mobile game "Pumpkin Garden", collects and analyzes patient behaviors within the game, generating reports for both patients and physicians to track medication responses and disease progression. However, there is still a lack of research on hypomimia digital therapeutic tools that can be widely used, and the implementation process, treatment strategy and evaluation criteria of digital therapeutic methods are not yet perfect.
  
  Parkinson's disease often leads to dysarthria \cite{dashtipour2018speech, melchionda2020perceptive, das2021automated}, which affects 89\% of patients and worsens as the disease progresses \cite{dashtipour2018speech}. In addition, cognitive impairment and mood disorders, especially depression, are the most common problems associated with Parkinson's disease (PD), placing a heavy burden on patients and caregivers \cite{litvan2012diagnostic, aarsland2021parkinson, jellinger2022morphological, pontone2019association, lintel2021mood}. Music therapy is an important form of digital therapy. Music can stimulate emotional expression, reduce stress and anxiety, and help patients relax \cite{ji2020daydream}. Music therapy is effective in improving the problem of speech disorders in Parkinson's disease, which can exercise and enhance the control of the throat and vocal cords. Patients with Parkinson's disease can try musical activities such as singing, chanting, or using musical instruments to work on their speech skills.LSVT-LOUD \cite{fox2012lsvt}, a Parkinson's disease-specific rehabilitation technique, is a high-intensity method of vocal training that effectively improves symptoms such as monotonous voice and low volume. Music influences mood and can support therapy. Barnish et al. \cite{barnish2020systematic} evaluated the potential benefits of a group-based performing arts intervention on quality of life, language, etc. for people with PD, and the results suggest that there are benefits of interventions such as music therapy, drama, and singing in the area of Parkinson's disease interventions.
  
  The aim of this study is to investigate the potential of digital therapies in reducing learning costs associated with rehabilitation of the hypomimia mask in Parkinson's disease and increasing patient initiative. The study will evaluate how interactive digital therapies can provide an engaging rehabilitation experience and create a positive atmosphere conducive to increasing patient self-efficacy.
\section{Formative Study}
The primary objectives are: (1) To elicit insights from neurologists regarding their perspectives on issues relevant to hypomimia. (2) To understand the patients' needs for developing a digital therapy system. In order to investigate how AI technology can be applied in the detection and rehabilitation of hypomimia patients with Parkinson's disease, we conducted a formative study to gather information on the following aspects: (1) Physicians' attitudes toward hypomimia treatment.(2) Patients' engagement in hypomimia rehabilitation. (3) Design opportunities for the AI interaction system.

\subsection{Method}
we proactively established formative study with neurologists and rehabilitation physicians in a hospital in China, aiming to gain a comprehensive understanding of the background knowledge. We collaborated closely with a group of four neurologists and two rehabilitation physicians. These physicians, each with an average of over ten years of experience in their respective fields, brought a substantial wealth of expertise to our research endeavor.

To explore the practical application of hypomimia therapy methods in rehabilitation, we conducted semi-structured interviews with rehabilitation professionals. We sought answers to the research questions. By addressing these questions: \textbf{RQ1}: How can artificial intelligence techniques be applied to help patients diagnose and treat Parkinson's hypomimia symptoms?
\textbf{RQ2}: How to design interactive tools for digital therapies applicable to patients with Parkinson's hypomimia symptoms? We aim to derive insights that can inform the design of AI-powered therapy systems within the HCI research community. At the same time, to study how treatment practices shape patient initiative and motivation.

\subsection{Key Findings and Design Implications}
\subsubsection{Cognitive problems in patients}
We found that the current diagnosis of hypomimia depends on the subjective expertise of neurologists, necessitating extensive clinical experience and requiring patients tow invest significant time and effort in hospitals. In the field of hypomimia rehabilitation, which usually involves verbal instructions from physicians, advising patients to engage in various facial exercises such as blowing up balloons, puffing cheeks, or blowing out candles to improve facial expressiveness. However, patients often lack a clear understanding of appropriate exercises and techniques and lack professional planning to guide them.

People with Parkinson's disease may have impairments in attention maintenance and information processing speed, which can affect their ability to multitask simultaneously. Interviews with physicians provided insight into the current state of digital therapies in healthcare organizations, which rely heavily on specific hardware setups. For example, there are eye-tracking games specifically designed to address cognitive disabilities. These interactive systems prioritize simplicity of user interaction and aim to be easy to learn and operate. However, facial movements are highly complex and there is a need to develop a tool to assess the complex dynamics of facial movements and provide feedback to patients. Combining the above issues, our goals are 1) Patients need a clear, intuitive, and user-friendly interaction interface that minimizes interaction complexity and reduces textual interfaces. 2) To conduct a rational rehabilitation program planning to present the rehabilitation process and effects in a quantitative manner.

\subsubsection{Emotional problems, such as depression and anxiety}
Through interviews conducted with neurologist, we found that patients with Parkinson's disease often experience various comorbidities, including emotional disturbances. Emotional issues such as depression and anxiety are common among individuals with Parkinson's disease, which may affect  patients' life quality and the disease progress. As a result of the facial expression disorder, patients will be less likely to look in the mirror and thus lack rehabilitation exercises for their hypomimia symptoms. Acceptance of the disease and the inability to get treatment in a timely manner can cause stress and anxiety for people with Parkinson's disease. Some have taken an approach that focuses on everyday life, allowing patients to value what they can do and find joy in [18, 20]. Relaxation activities that encourage patients to be in a calmer state of mind have likewise been described as beneficial. Combining the above issues, our goals are 1) the need to activate patients' initiative and motivation to participate in the rehabilitation process with ease and pleasure. 2) the use of art forms, such as music therapy, to elevate the patient's mood and reduce anxiety and depression. Rhythmic activities are used to help patients relax their faces and reduce muscle stiffness and tension.

\subsubsection{Lack of active rehabilitation awareness}
Hypomimia may be difficult to detect in the early stages because its onset is usually gradual, resulting in a possible delay in awareness of the problem by the patient, family, and physician. Patients should be actively involved in the rehabilitation program, recognizing that rehabilitation is a long-term process that needs to be sustained and should not be interrupted even when symptoms improve. Interviews with rehabilitation physicians revealed that recovery from hypomania relies on passive therapeutic interventions such as manual massage, electrical stimulation, and electrotherapy. Active treatment requires one-on-one guidance from a rehabilitation therapist. It works better than passive stimulation methods if the patient actively participates in sensory input, processes information, observes his or her facial movements, and receives immediate feedback, creating a closed-loop process. However, due to its high cost and the shortage of hospital-based rehabilitation therapists, active therapy has not been widely used clinically [citation needed]. Based on these observations, we aim to 1) establish an accessible digital detection and treatment system to provide targeted guidance to patients. 2) activate patients' agency and motivation through active training content, thus allowing patients to have a sense of autonomy in their rehabilitation.

\subsubsection{Continuous monitoring of patients' symptoms}
The severity of hypomimia is intrinsically associated with disease progression and individual variances. Typically, its severity exacerbates as the disease advances; in the early stages of Parkinson's disease, facial symptoms may be relatively mild, but they tend to intensify over time. Each Parkinson's disease patient follows a distinctive clinical trajectory, with some experiencing greater facial stiffness while others may be less affected, leading to varying levels of impairment in their facial expressive capabilities. For people with Parkinson's disease, recording symptoms or learning to self-monitor is an important part of self-management. One study found that participants used a “tracking log” for self-monitoring and reported that this improved healthcare interactions and provided a basis for discussion during counseling [21]. Patients need regular medical review and ongoing medical support to make timely adjustments to their treatment regimen. Our goals were 1) to effectively manage and analyze the data collected by ai in order to provide better treatment recommendations to patients. 2) to provide timely feedback on problems and progress during rehabilitation so that therapists can adjust treatment strategies.

\subsubsection{Classify actions and difficulties}
Individuals with Parkinson's disease have individual differences in symptoms, with each person having a different type of symptom and degree of disease. It is especially important to customize the information and match it to the stage of the disease. Effective rehabilitation for hypomimia requires comprehensive coverage of various facial regions to restore facial muscle flexibility and motor control. Additionally, facial muscle rigidity and motor control issues may extend to the muscles in the head and neck, including those involved in throat and tongue function, potentially negatively impacting articulation and swallowing. Consequently, our design principles aim to create a comprehensive and professional rehabilitation training guidance method while categorizing various facial actions and levels of difficulty.


\subsection{Design Goals}
Based on previous research and our findings, we propose three main design goals: 1) A professional and quantitative digital training program based on anatomy and rehabilitation. 2) Interactive interfaces with timely and objective feedback for individuals with masked face symptoms. 3) Continuous patient involvement to enhance patient initiative and self-efficacy.

\section{System Design and Implementation}
\subsection{Hypomimia Detection}
\begin{figure*}[ht]
    \centering
    \includegraphics[width=0.8\linewidth]{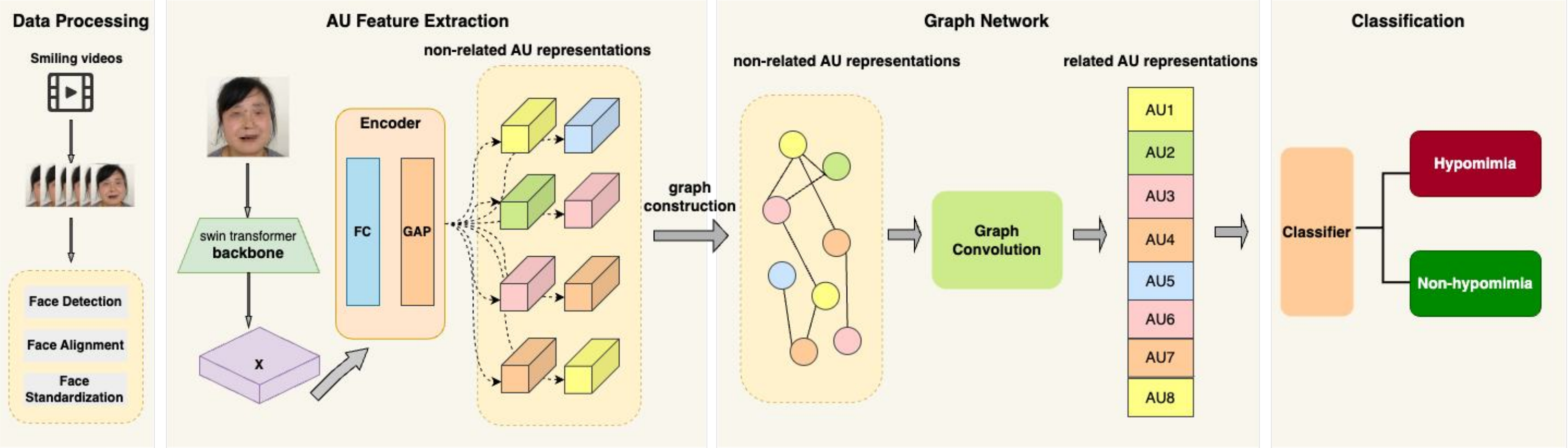}
    \caption{The pipeline of the proposed method, comprises four essential steps: data collection and preprocessing, AU feature extraction, AU feature fusion, and hypomimia classification.}
    \label{fig:method}
\end{figure*}

The detection pipeline includes 4 steps, which are data collection and preprocessing, facial AU feature extration, facial AU  feature fusion and hypomimia classification. Figure 2 \ref{fig:method} depicts the pipeline of the proposed method. Hypomimia in Parkinson's disease, characterized by reduced facial movement and expression, can be more precisely captured through the analysis of Action Units (AUs). Graph Convolutional Networks (GCNs) further enhance model performance by leveraging the interconnectedness of AUs.

\subsubsection{Data collection and preprocessing} 

Due to the limited public dataset in hypomimia, we collected hypomimia dataset to train the detection model. The dataset was collected at a hospital's Department of Neurology, ensuring that all procedures involving human subjects adhered to established ethical guidelines. The total hypomimia videos is 105 sample,  consisting of 55 facial video recordings from healthy subjects and 50 from patients with Parkinson's Disease. We recorded frontal facial movements of participants using a smartphone in 1920×1080 pixels with 60 frames per second. Participants were first asked to face the camera with a neutral expression, followed by a gentle smile. The video captured both the neutral and smiling expressions, with approximately 30 seconds of footage collected from each participant. Stringent controls were implemented for variables including lighting conditions and head positioning.

In the data preprocessing, the MTCNN[48] model was utilized for frame-by-frame segmentation of the video streams, enabling facial recognition, alignment, and normalization for each frame. Following MTCNN processing, the smiling videos were partitioned into training, validation, and test sets based on individual participants, employing the hold-out method. This resulted in a 60-20-20 split, with 60 participants in the training set, 20 in the validation set, and 20 in the test set. The corresponding number of video frames for each set was 20,479, 6,898, and 8,030, respectively.

\subsubsection{Facial AU feature extraction} 

After processing each frame of the facial images, a pretrained Swin Transformer[49] model is utilized to produce a 1x2048 feature vector encapsulating facial information. The encoder encompasses both fully connected (FC) layers and global average pooling (GAP) layers. Within the FC layers, eight AU extractors were responsible for feature extraction from facial representations, resulting in eight distinct AU feature maps. The GAP layer computed an average value for all pixels in each channel. Subsequent to this global average pooling step, weed obtain eight separate feature vectors. 
Facial Action Units (AUs) are not independent features but rather an interconnected system. Based on this characteristic, we employ a graph neural network (GNN) to aggregate the relationships between facial features. The acquired  AU features were considered as nodes in a graph. Relationships  considered as edges among AUs were established using k-nearest neighbor (k=2) connectivity, determined by node proximity.  Subsequently, after constructing the graph structure of AUs, graph convolution operations  were employed to convert unrelated AUs into related AUs.

We adopted a transfer learning approach to enhance AU feature extraction. Initially, the model was pre-trained on a dataset with AU labels, focusing on learning the features associated with eight specific AUs: AU1, AU2, AU4, AU6, AU9, AU12, AU25, AU26 from the DISFA dataset \cite{mavadati2013disfa}. By leveraging the pre-trained model's ability to extract fundamental facial features, this approach mitigates the challenges posed by the scarcity of hypomimia dataset.

\subsubsection{Facial AU feature classification} 

After graph convolution, we acquired eight AU features containing their relational information. The eight features were used as inputs for training a neural network (NN), comprising two convolutional layers (CNN) \cite{lecun2015deep}. Stochastic gradient descent(SGD) was employed as the feature optimizer, and the cross-entropy loss function served as the loss function. The results in Table \ref{tab:results} illustrated the discerning efficacy of this algorithm, highlighting its superior discrimination capability compared to other machine learning models, thereby validating the efficacy of utilizing AU features for masked face identification.

\begin{table}[htbp]
\centering
\caption{Results (in\%)on validation and test sets}
\begin{tabular}{ccccc}
\toprule
Model & Accuracy & PPV & TPR & F1-score \\
\midrule
RF & 84.5 & 84.8 & 82.7 & 83.7 \\
SVM & 88.7 & 88.0 & 88.7 & 88.3 \\
Our method & 91.7 & 92.8 & 90.6 & 91.7 \\
\bottomrule
\end{tabular}
\label{tab:results}
\end{table}

\subsection{Hypomimia Rehabilitation}
\subsubsection{AU-based rehabilitation training design} 
Building upon a thorough review of existing facial rehabilitation techniques for hypomimia, we introduce a novel, interactive training system that leverages Action Units (AUs). By integrating targeted exercises for the eyebrows, ocular muscles, cheeks, nose, lips, and articulation with AU-based feedback, we have developed a personalized rehabilitation approach that empowers users to actively engage in their recovery process. Table \ref{table:facial_training}(Appendix) presents a detailed overview of our AU-based rehabilitation trainings.

\begin{table*}[htbp]
\centering
\caption{Facial expression training details}
\label{table:facial_training}
\begin{tabularx}{\textwidth}{XXXX}
\toprule
Facial area & Eyebrow & Nose and eye & Lip \\
\midrule
Training method & Frown exercise, eyebrow raising exercise & Hard blink and nose wrinkle & Smile and grin practice \\
\midrule
Training content & 
  \parbox[t]{0.23\textwidth}{Practice frowning and raising eyebrows repeatedly, with patient instruction to center both eyebrows and perform forehead surprise} &
  \parbox[t]{0.23\textwidth}{Continuous eye opening and closing, forceful eye opening and closing, and nose stretching upward} &
  \parbox[t]{0.23\textwidth}{Completion of pout, smile, toothy smile, and other consistent movements, along with patient guidance for outward and upward corner-of-mouth motion} \\
\midrule
AU coding number & 
  \parbox[t]{0.23\textwidth}{AU1: Inner Eyebrow Elevation\\ AU4: Eyebrow Lowering and Forceful Furrow} &
  \parbox[t]{0.23\textwidth}{AU5: Forceful upper eyelid elevation and intense stare\\ AU44: Forceful blinking and eye closure\\ AU9: Nose wrinkling} &
  \parbox[t]{0.23\textwidth}{AU12: Mouth Corner Stretching\\ AU13: Cheek Puffing\\ AU19: Tongue Movement\\ AU27: Mouth Widening\\ AU24: Lip Pursing\\ AU18: Lip Pouting\\ AU33: Cheek Inflation for 3 Seconds\\ AU28: Lip Suction} \\
\bottomrule
\end{tabularx}
\end{table*}

\begin{figure}[ht!]
    \centering
    \includegraphics[width=0.99\linewidth]{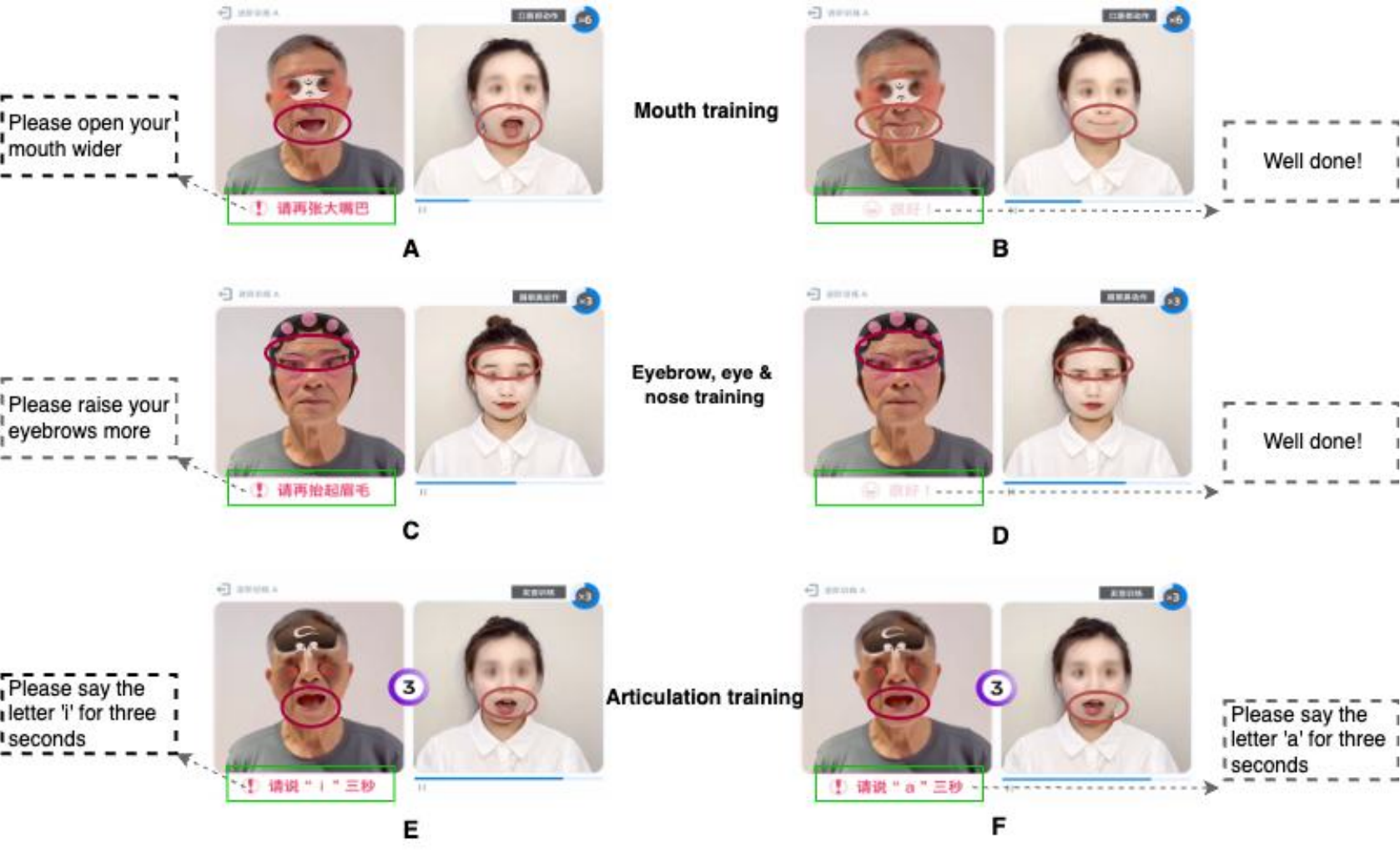}
    \caption{Screenshots of HypomimiaCoach: A) improvement feedback in mouth training, B) fine feedback in mouth training, C) improvement feedback in eyebrow, eye and nose training, D) fine feedback in eyebrow, eye and nose training, E) improvement feedback in articulation training, F) fine feedback in articulation training.}
    \label{fig:interaction}
\end{figure}

To recognize AUs, we leveraged the openGraphAU \cite{luo2022learning} model, which excels at identifying the movements associated with 41 distinct AUs and quantifying their activation levels. This model provides a quantitative measure of facial movements, enabling us to precisely track changes in facial expressions over time. As depicted in Fig. \ref{fig:interaction}, HypomimiaCoach employs this model to analyze user facial expressions and offer tailored feedback.

\subsubsection{Chinese-opera based music interactive design} 
Music therapy, which involves regular sessions with a qualified music therapist, has been shown to improve mood through emotional expression \cite{aalbers2017music}. Traditional Chinese opera, with its centuries-old history, unique charm, and profound cultural impact \cite{wu2019determinants}, is characterized by diverse regional variations and evokes empathetic responses from local audiences. Recognizing the potential benefits of Chinese opera's phonetic distinctiveness for improving single-character pronunciation training in seniors, we have incorporated it into our rehabilitation program.

To enhance the effectiveness of the training, we strategically matched different training content with specific styles of traditional Chinese opera. For instance, lively operas like "Rolling Lantern" are well-suited for lip movement exercises, while melodious, extended pieces like "The Case of Chen Shimei" are ideal for pronunciation practice. To further engage users and make the training experience more enjoyable, we integrated opera-themed filters tailored to each opera's style (Table \ref{tab:results} and Fig. \ref{fig11} in Appendix). These filters, activated during transitions between operatic segments, create an interactive "face-changing" effect, enhancing the overall training experience.

\subsubsection{Basic and advanced training mode} 

We offer two modes of rehabilitation training: basic training and advanced training. Fig. \ref{fig:mode} in Appendix outlines the procedure for both.

Basic training comprises exercises targeting distinct facial regions, including the eyes, brows, nose, lips, swallowing muscles, and articulation. This training follows an iterative cycle of instruction, exercise, and feedback. Users begin by selecting a training session. The system then presents guided facial movement demonstrations. Following this instructional phase, users perform the designated movements. An AI model analyzes users' facial expressions in real-time, assessing the accuracy of their movements and providing immediate on-screen feedback. Users can adjust their expressions based on this feedback and repeat the exercise until they achieve the desired outcome. Advanced training involves facial operatic exercises, allowing users to choose training content based on their preferences for duration and difficulty. Upon entering the game, users synchronize their facial movements with instructor guidance. The AU recognition model continuously analyzes facial movements, identifying and evaluating the accuracy and deviations of user actions. The system provides visual feedback categorized into three levels: "perfect," "good," and "come on."
A personalized facial rehabilitation training feedback mechanism has been developed to cater to the diverse facial conditions of different patients. The system provides real-time feedback and an overall performance score upon completion of each training session. The system evaluates the accuracy of a user's facial movements by calculating the discrepancy between their current expression and a neutral baseline expression within specific facial action units. There are distinct user interfaces designated for patients and physicians. The physician's interface is equipped with the capability to aggregate data from each instance of patient training, thereby enabling a comprehensive assessment of the conditions of various facial regions. This functionality facilitates a more informed and targeted approach to treatment and monitoring.

\section{User Study}
To evaluate the system, we conducted a user study focusing on participants' engagement, initiative, and self-efficacy during rehabilitation training. The study was conducted at a public hospital in China and received ethical approval.

\subsection{Participants}
We conducted two evaluation experiments: the first, patient-oriented (n=7 Parkinson's Disease patients with hypomimia, average age: 69.57, average MDS-UPDRS hypomimia severity score: 2.29); the second, physician-focused (n=10 physicians, 7 from neurology department, and 3 from rehabilitation department). The demographic characteristics of the participants are reported in Table \ref{tab:participants}.

\begin{table}[htbp]
\centering
\caption{Demographics of the participants in the user study}
\begin{tabular}{cccccc}
\toprule
ID & Gender & Age & \begin{tabular}[c]{@{}c@{}}PD duration \\(years)\end{tabular} & \begin{tabular}[c]{@{}c@{}}Hypomimia \\ Severity\end{tabular} & \begin{tabular}[c]{@{}c@{}}Depression \\ Severity \end{tabular}\\
\midrule
PD1 & Male & 71 & 3 & 3 & 2 \\ 
PD2 & Male & 63 & 1 & 1 & 3 \\ 
PD3 & Male & 65 & 16 & 2 & 5 \\ 
PD4 & Male & 80 & 3 & 3 & 3 \\ 
PD5 & Male & 72 & 1 & 0 & 2 \\ 
PD6 & Male & 62 & 4 & 2 & 3 \\ 
PD7 & Female & 74 & 3 & 5 & 4 \\
\bottomrule
\end{tabular}
\label{tab:participants}
\end{table}

\subsection{Procedure}
\subsubsection{Experiment Setup}
This study was carried out in a dedicated communication room within the neurology ward of a hospital, which consisted of two distinct sections: an interview area and an experimental area. In the interview area, a research facilitator conducted participant interviews, while a recorder was responsible for audio recording and documentation, with both participants and their caregivers present. In the experimental area, an operator assisted users with system usage, a physician observed participants' interactions with the system, and another recorder was responsible for observing, and recording video footage, with once again both participants and their caregivers present. 

\subsubsection{Pre-test}
The project's content and objectives were introduced to participants and their caregivers. Ethical concerns regarding data usage and privacy protection were clarified. After obtaining consent from participants and their caregivers, informed consent forms were signed. Prior to the start of the study, each participant underwent an initial assessment that included a facial expression test, a depressed mood test, and a self-perception of hypomimia symptoms. An interview was conducted to ask the participant about basic information, symptoms, duration of onset, pre- and post-onset differences, and acceptance of electronic devices. A physician assessed the severity of the participant's hypomimia. Participants were introduced to the rehabilitation training system and received brief training on how to use it. They were then asked to use the system for 20 to 30 minutes of supervised rehabilitation session.

\subsubsection{During Test}
Participants were invited to the experimental area and seated in computer chairs facing computer screens. Initially, the operator guided participants in assessing hypomimia using a facial detection model. Subsequently, rehabilitation training commenced. The operator provided guidance on using a one-minute version of our system for facial training. After participants were familiarised with the use of the system, a round of 3-minute advanced training and a round of basic training for each area of the face were performed. A doctor and caregiver were present throughout the experiment to observe the participants' behaviour.

\subsubsection{Post Test}
After testing, participants and their caregivers were invited back to the interview area for a 15-minute semi-structured interview. The questionnaire was presented in the form of a modified interview with the participant in order to accommodate potential cognitive challenges for some participants. The questions included an assessment of system usability, system engagement and self-efficacy. Further interviews were conducted to address the participant's experience of using the system. After the participants' experiment, physicians were tasked with completing a project-specific questionnaire. While completing the questionnaire, physicians were encouraged to offer additional explanations for their ratings on specific items. Following this, semi-structured interviews were conducted with the physicians to delve further into the questionnaire items and gain more profound insights into their feedback and perspectives.

\begin{figure}[ht!]
    \centering
    \includegraphics[width=0.99\linewidth]{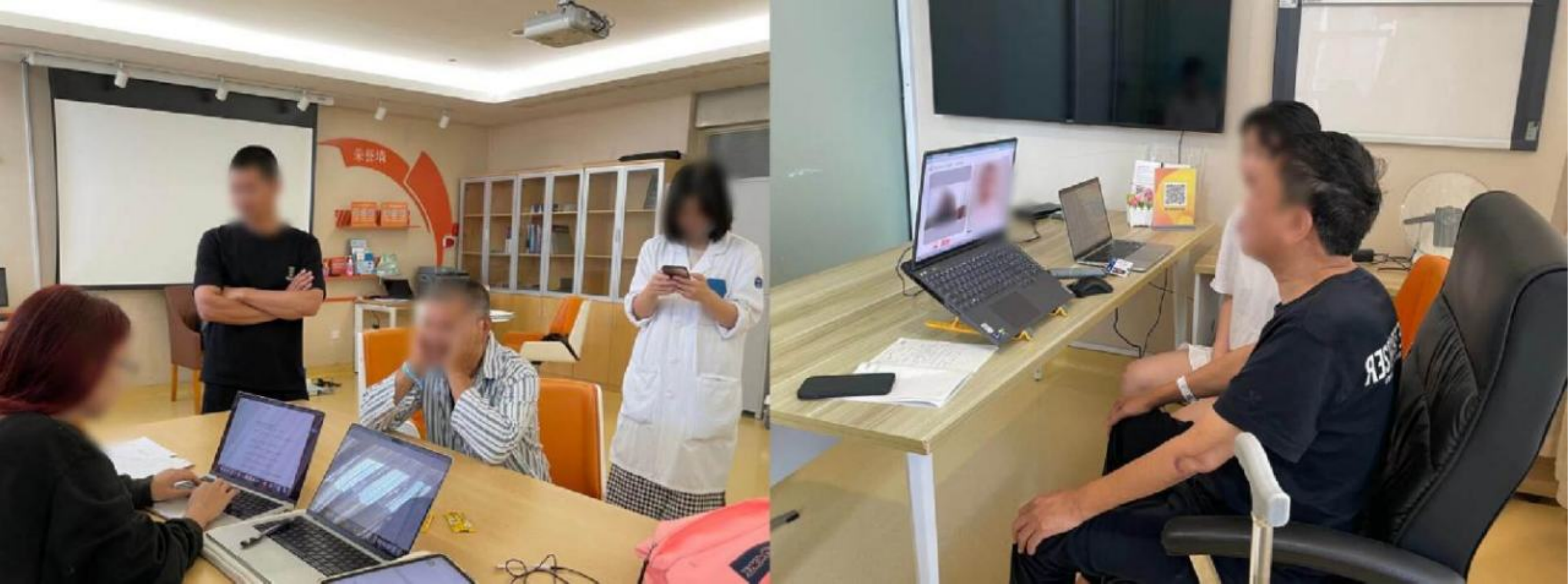}
    \caption{A participant interacting with HypomimiaCoach}
    \label{fig:mode}
\end{figure}

\subsection{Measures}
\subsubsection{Assessment of engagement and autonomy}  
As our formative survey showed, increasing participants engagement and autonomy in their rehabilitation training activities was a key design goal of our system.
Understanding user preferences and satisfaction with the system by observing the duration and level of engagement of users interacting with the system can help assess willingness and level of participation.
Autonomy can be indirectly assessed through questionnaires that allow subjects to rate their level of efficacy in specific tasks or domains. In addition, self-efficacy is associated with a reduction in depressive symptoms. Thus, by comparing participants' mood state assessments before and after using the system, we can further reveal changes in self-efficacy.

\subsubsection{Usability Assessment}
In order to navigate each user flow, participants had to make specific facial expression movements or articulations. Therefore, to assess participants' successful application of rehabilitation training maneuvers during the system use flow, we captured the number and type of errors they made in the training content (e.g., task comprehension errors or invalid input entries), as well as the participants' efficiency in obtaining help from the physician and research assistant. In addition, participants' success in the assigned tasks was related to rehabilitation ability; therefore, we used this usability assessment at the end of users' use of the system. 

\subsubsection{Interactive performance} 
The aim of this study was to assess participants' performance in interacting with the Parkinson's disease hypomimia rehabilitation training system. We paid particular attention to clinical practice issues with the user interface and the potential impact of the system on the user's rehabilitation process. We observed the extent of participants' hypomimia symptoms as measured by accuracy in completing tasks during this process. We analyzed participants' duration of each round of training tasks and the types of exercises completed. Accuracy refers to the degree to which the movement meets the standard, and we assessed the accuracy of the participant's movements using au values, which were categorized into three levels. Finally, participants' feedback was collected through open-ended questions to understand their subjective experience of the system.

\subsubsection{Semi-structured interview} 
To gain insight into the user experience of the Parkinson's Disease (PD) Facial Expression Rehabilitation System, we conducted semi-structured interviews with participants to explore users' perceptions, experiences, and expectations of the system in a flexible and direct manner. We interviewed medical professionals, including physicians and rehabilitation therapists, to assess the system's therapeutic approach, functionality, and potential impact on PD participants with facial hypokinesia. Physicians were asked about their recommendations for using the system in rehabilitation therapy and potential problems that might arise. Each interview was recorded by consent and transcribed for detailed analysis.
\section{Results}
\subsection{Engagement of Participants}
\subsubsection{\textbf{Task completion}}
All participants successfully completed both rounds of the rehabilitation training tasks, the average length of the basic and advanced training was three minutes, and all participants successfully completed each task, with overall accuracy varying from person to person. Almost all participants understood the training content and participated attentively. In the lip and articulation portion of the training, participants had lower mean accuracy scores, and there were instances where lip movements were misunderstood and movements were not performed properly. The articulation training most participants understood the articulation movements and blended well with the music and with the opera background. The majority of participants felt that the system was definitely helpful for pronunciation improvement.
Feedback was collected from users on the difficulty of the tasks and their participation experience. Some participants found the task challenging, \textit{"Simple movements are good to learn. Combinations of the more complicated movements felt bad to do and slower would be better. The basic training model was relatively easy and a little easier to imitate."}

For example, PD3 is an individual experiencing early-stage symptoms of Parkinson's disease. He is a former enthusiastic singer, has faced recent challenges in facial and vocal expressions, resulting in unclear speech and emotional distress. He stated, \textit{"I noticed that I sounded weaker compared to before, which made me reluctant to sing. My doctor advised me to work on my face and talk more, but I didn't know what to do."} 
After using the system, pd3 felt that this game form of rehab was better than traditional rehab, but felt that it was a little complicated to use for the first time, \textit{ 'I didn't know the difference between the two modes when I used the system, it felt like both were similar because I didn't get into the groove."}

PD7, another participant with early-stage Parkinson's symptoms, was previously an outgoing individual who enjoyed making dance videos on TikTok. The disease, however, led her to feel insecure and anticipate a decline in her quality of life. She experienced daily mood fluctuations. After trying our system, she found the tasks to be highly correlated with her rehabilitation goals, stating, \textit{"I find this format wonderful. Music synchronized with facial exercises significantly improves my mood. The exercises make me feel that my symptoms have improved. The physician recommended daily exercise, and with a system like this, I can engage in systematic training.  I believe that through these exercises, I can delay the potential onset of symptoms and slow down the progression of the disease."}

\subsubsection{\textbf{Participants' satisfaction interacting with the system}}
Understanding user satisfaction with the system can help assess willingness and level of participation.
A 5-point Likert scale was used to collect participant and physician ratings of satisfaction with the system's interaction and digital therapy. In terms of interaction, participants' mean scores for satisfaction with the interaction method (M = 3.29, SD = 0.88), satisfaction with the opera music (M = 3.57, SD = 0.73), and satisfaction with the operatic filter (M = 3.33, SD = 0.94). Through questionnaires and semi-structured interviews, most participants perceived the music format as engaging and encouraging to sing it out together. This preference suggests that systematic digital therapy is attractive and effective for these participants, improving their self-efficacy by making the recovery process more enjoyable and manageable.

However, satisfaction with Chinese opera music mainly depended on the individual participant's interest in opera music. Participants rated the opera filter moderately, believing that they did not enhance the entertainment value of the training exercise. Some expressed the opinion, \textit{"I didn't notice the opera filters."} and \textit{"I don't care if there are operatic filters as long as the rehab relieves the symptoms of hypomimia."}
Of the 7 participants, 6 participants reported feeling relief in their face after completing the training, 5 participants felt that the method improved their articulation, and all 7 participants reported feeling more emotionally relaxed. PD7 thought the system was helpful for mood improvement, \textit{"I feel curious about the system. Feeling in a good mood because it feels helpful. If my doctor requires that I train every day, then it's okay for me to train once or twice a day."}

\subsection{Initiative of Participants}
Through the questionnaire survey of the participants found that like to use (M = 4.43, SD = 0.90), would like to use frequently (M = 4.29, SD = 0.70). These scores indicate that participants found the system engaging and useful, which is a strong indicator of self-efficacy as it reflects their confidence in using the system to meet their rehabilitation needs. Among the 7 participants in the experiment, virtually all participants reported having no prior experience with digital therapeutic modalities. Of these, 6 participants found this approach convenient and expressed their willingness to adopt it. Participants exhibiting Hypomimia particularly appreciated our system and expressed a desire to take it home for personal use, citing its convenience. PD5 said, \textit{"If there is really an improvement , I would like to use it, it will help me with my illness. I can do rehab when I have nothing to do at home."} 

5 participants thought that this game form of rehabilitation was better than the traditional form of rehabilitation, and some participants thought that tai chi whole body exercise was better.
PD6 agreed that, \textit{"Comparatively, it's more convenient. After falling ill, the physician suggested practicing the 'eight pieces of brocade (a set of qigong exercises with assumed medical benefits)', but I never attempted it because I had no strength in my entire body. However, with this format, I feel I can engage in exercise."} This willingness to embrace new approaches, especially after limited exposure, indicates high levels of self-efficacy as it demonstrates confidence in their ability to learn and benefit from the system.

From interviews with rehabilitation physicians, at present, treatment in the rehabilitation department takes a more passive form, because active massage often requires more labour costs and higher treatment prices, so the treatment generally carried out in the rehabilitation department is passive. The immediate feedback received during active training enhances motor learning and functional recovery. This form of active training is very helpful for rehab training. Physicians' perspectives confirm the system's potential to improve self-efficacy by providing an effective, positive approach to recovery. A rehabilitation physician said in an interview that,

\textit{"There's a scarcity of specialized software like this, and it's genuinely needed. Typically, patients are instructed verbally on how to train. However, in terms of effectiveness, active training is relatively better because there is a closed loop from input to output.
This is attributed to the active engagement of participants in initiating and modulating movements."}\

\subsection{The Appropriateness of the System}
To assess the usability of the system, we collected feedback from participants and physicians. Participants completed a SUS usability questionnaire on a scale of 1-5, including ease of use (M = 3.50, SD = 1.12), ease of learning (M = 4.67, SD = 0.47), comfort (M = 4.00, SD = 0.71), ease of recall (M = 4.67, SD = 0.47), fun (M = 3.29, SD = 0.88). These results indicate that the system was designed to effectively meet the target population's usability needs and provide a clear and intuitive user experience.
The physician also completed a usability questionnaire on a scale of 1-5, including fault tolerance (M = 3.88, SD = 0.78), suitability for clinical use (M = 4.00, SD = 0.87), effectiveness (M = 4.29, SD = 0.70), timely feedback (M = 3.71, SD = 0.70), the usefulness of basic training (individual exercises for different parts of the face) (M = 3.50, SD = 0.87) , the usefulness of advanced training (integrated exercises for different parts of the face) (M= = 3.88, SD = 0.93). Clinical interviews with physicians further revealed the clinical feasibility and value of the system. Neurologists commented positively on the system's smoothness and enjoyability, noting that participants seemed to have a positive experience.
The consensus among both participants and physicians was that the system possessed favorable usability. In the interviews, neurologists made comments such as: \textit{"It's quite smooth."},\textit{ "It's rather enjoyable; the first participant seemed to have a good time."} The feedback on usability indicated a consensus among both physicians and participants regarding the favorable usability of the system. Suggestions for improvement included incorporating more positive feedback to enhance participants motivation.

\textit{"Adding some positive feedback could be beneficial to encourage patients. For example, after completing an exercise, providing feedback like 'Congrats, you've taken another step towards better health' can boost patient motivation. These comments reflect the generally positive sentiment about usability, along with helpful suggestions for improvement."}

Moreover, rehabilitation specialists highlighted the importance of active participant engagement in the recovery process. Unlike passive modalities commonly used in rehabilitation, active training, such as facial exercises, has been shown to yield better outcomes. This system aligns with this principle by encouraging participants to participate actively in rehabilitation.

\textit{"This system guides patients in performing active exercises for their orofacial muscles. Rehabilitation inherently emphasizes patient engagement. Orofacial exercises are essential, and ideally should be widely adopted among Parkinson's patients, much like the daily use of antihypertensive medication for hypertension. All Parkinson's patients should be equipped to perform these exercises regularly and comprehensively. "}

Physicians emphasized the potential of this system to slow disease progression and promote participant adherence. One neurologist suggested a progressive training approach, starting with isolated muscle groups and gradually progressing to more complex facial movements and functional tasks.

\textit{"A progressive approach, starting from isolated muscle groups to integrated facial movements, is recommended. For instance, training can begin with oral exercises, followed by facial and periorbital exercises. Subsequently, the focus can shift to coordinated movements of the entire face, such as expressing various emotions. Finally, exercises can be integrated with functional tasks, like engaging in conversations with therapists or performing expressive reading."}
\section{Discussion}
In this section, we discuss the potential of this system for use in future clinical practice and provide a set of insights for future designers entering the domain.
 \subsection{Reasonable Training Time, Frequency and Practice Strategy}
    In this study, we gain insights into the timing and methodology of rehabilitation training. This combined detection and rehabilitation system can adjust and optimize treatment plans to be more targeted based on patient-specific conditions and feedback. Artificial intelligence algorithms can detect Parkinson's disease by analyzing a patient's facial expressions and movements, an approach that improves diagnostic consistency and accuracy. Digital therapies can help doctors and patients understand the condition and recovery process more objectively. This quantitative assessment avoids the subjectivity of traditional scale assessments and allows for more accurate monitoring of rehabilitation outcomes. One participant suggested shorter training sessions, stating, \textit{"It's best not to make each session too lengthy; sitting for extended periods isn't beneficial for the lower back."} Rehabilitation physicians provided instructions on the duration and approach to training, , and suggested the feasibility of using the system for rehabilitation training at home,\textit{"For example, within a hospital setting, begin by instructing patients on how to use the system and then encourage them to use it in their homes for long periods of time. Frequency is 5 times a week with weekends off, 1-2 sessions per day, each lasting 20 minutes."}
    
    A rehabilitation physician commented, \textit{"The process should begin with localized training and progress to overall training, from muscle-specific exercises to functional exercises. For instance, initiate with oral exercises, then transition to facial exercises, and subsequently target the areas around the eyes. After addressing individual areas, integrate the entire face into coordinated exercises, such as performing various expressions by engaging multiple muscle areas synergistically."}

Compared to patients with less severe symptoms of Parkinson's disease, those with more severe symptoms, which are often accompanied by poorer cognitive performance and lower health status, have a relatively low need for treatment for hypomimia, a symptom of Parkinson's disease that manifests itself as a reduction in facial expression. Early detection and intervention for patients exhibiting early symptoms of Parkinson's disease makes a lot of sense. This is because in the early stages of the disease, patients have relatively better cognitive function and overall health, which makes them better able to respond to treatment and thus more likely to benefit from treatments for the relief of motor symptoms such as hypomimia. Therefore, focusing on early recognition and intervention for hypomimia is important for improving treatment outcomes and patient quality of life. These findings indicate the potential of HypomimiaCoach system and the need for further research on their integration with traditional treatments. 

\subsection{System Interaction Experience}
\subsubsection{\textbf{Further explanation of complex and ambiguous movements}}
In this study, we observed that unimodal interaction systems perform poorly in user engagement, particularly for patients with Parkinson’s disease. Due to potential cognitive limitations, these patients often struggle to accurately comprehend or respond to complex verbal instructions \cite{jian2014modality}. Consequently, we recommend that the design of multimodal dialogue interfaces specifically consider the cognitive abilities of users, ensuring that the interaction modalities are both intuitive and easily comprehensible.

To this end, providing visual and auditory feedback is crucial, utilizing images, videos, or audio cues to support verbal instructions \cite{yu2011interactive}, thereby aiding patients in understanding the required actions. Furthermore, the supervision and guidance of healthcare professionals during system use are essential, as they can offer necessary support and interventions \cite{hennig2012developing}.

By implementing multimodal dialogue interfaces, we aim to enhance the effectiveness of rehabilitation training for patients, improve user interaction experiences, and increase the feasibility of these systems within clinical settings. However, challenges remain in voice guidance, particularly regarding dialectal variations, necessitating explanations from caregivers, which underscores the need for further research in the context of artificial intelligence systems. To ensure the requisite intelligence of such systems, it is critical to involve a broader range of stakeholders. Thus, raising awareness and knowledge of digital therapies and rehabilitation among caregivers (such as family members) is imperative, enabling physicians to tailor training plans based on the patient’s disease progression while caregivers assist patients in executing these plans \cite{dharma2018increase}.

Through experimental observations, we have identified instances where patients may misinterpret certain gestures made by the instructor. For example, during tongue exercises, when the instructor points to their lips with an index finger, some patients tend to mimic the finger pointing to the lips without actively engaging in the required tongue movements. This highlights certain limitations in the system's human-computer interaction capabilities. In situations involving complex or ambiguous actions, the addition of voice explanations becomes essential to assist users in understanding the fundamental aspects of the exercises.

\subsubsection{\textbf{Further intensification of pronunciation training}}
During the experiment, participants demonstrated a strong willingness to engage in pronunciation training, and both participants and physicians generally acknowledged the significant effectiveness of pronunciation training in improving articulation and facial muscle effects. However, the current system has only incorporated basic vocal exercises and lacks specialized vocal training, indicating the need for further development in this area. Recent research on computer-assisted pronunciation training (CAPT) has explored various approaches to improve learner outcomes \cite{minematsu2008training, cheng2020exploiting}. 

The proposed system aims to offer a comprehensive assessment of articulation functions by integrating both subjective and objective measures. This assessment is envisioned to include evaluations of the articulatory organs, motor functions, pronunciation, and conversational skills. It is anticipated that the system will analyze various phonetic parameters, such as jaw distance, tongue distance, tongue area distance, oral alternation movement rate, onset time of voiced sounds, duration of phonetic features, trends, airflow time ratio, ratio of voiced to voiceless sounds, speech type, and articulation clarity, along with generating relevant tongue position maps and voice position diagrams \cite{ziwei2014multiparameter, si2015automatic}.

In addressing abnormal articulation, the system is designed to incorporate five targeted training modules: articulation correction, articulatory organ movement, articulation motion training, speech training, and tongue twister exercises. These modules are planned to feature a variety of engaging rehabilitation activities, with differing levels of difficulty and sensitivity to cater to individual patient needs. Furthermore, the system intends to include voice library management capabilities, which will facilitate the collection and management of phonetic parameters from diverse dialectal populations \cite{zhou2009introduction, igras2014acoustic}, thereby enhancing the accuracy of assessments and the precision of training interventions.

\subsection{The Application Strategy of Digital Therapy in Rehabilitation Training}
\subsubsection{\textbf{Inspiration in rehabilitation training of facial disorders}} 
Early detection through artificial intelligence algorithms allows for timely rehabilitation training and slowing of disease progression, which is of great significance in early intervention and prevention of hypomimia. Visual feedback through the system helps to improve the patient's knowledge and management of the disease. In addition digital therapies combined with artificial intelligence make remote rehabilitation possible, which is especially beneficial for patients living in remote areas or with limited mobility. Tele-rehabilitation can reduce the uneven geographical distribution of healthcare resources and enable more patients to receive specialized rehabilitation. 

In the long term, this approach can be used to intervene and train individuals with facial expression disorders such as stroke or facial paralysis \cite{moosaei2019modeling}. It can integrate theories of facial rehabilitation for other facial abnormalities, design corresponding active facial rehabilitation exercises, and combine them with AU recognition models for rehabilitation training. HypomimiaCoach underscores the versatility of digital rehabilitation therapy across medical domains, opening new avenues for future research and treatment. It has the potential to offer effective rehabilitation and support to a broader patient population. This paper provides ideas for the treatment of other facial diseases. The application of intelligent rehabilitation systems may drive a shift in the traditional healthcare model to a more patient-centered and technology-assisted model.

\subsubsection{\textbf{Different preferences in Chinese opera music and filters}}
In this study, we focused exclusively on Chinese opera music. Our findings indicate that patients do not demonstrate a strong preference for this genre and show limited interest in opera filters. Moving forward, it would be beneficial to explore a wider variety of music genres to accommodate the diverse musical tastes and preferences of patients. 
Incorporating the rhythmic elements of opera music into exercise regimens may enhance patients' familiarity, relevance, and motivation \cite{ziv2011music}. Interviews with patients revealed that while preferences did not significantly impact their engagement, there was a high level of acceptance for the music used. Preferences for music types may vary regionally, as different areas in China are associated with distinct styles of opera \cite{ccift2020effect}. Additionally, considering Self-Determination Theory, we anticipate that personalized music conditions will yield higher intention to use, as patients will have greater autonomy in selecting background music that they find more relevant to their experiences \cite{heiderscheit2014music}. These findings enhance our understanding of how music can be effectively integrated into learning environments and underscore the importance of considering cultural and individual preferences in music selection.

\subsection{Ethical Considerations}
When handling patient facial images, techniques such as "Digital Mask" should be employed to anonymize patient identities while preserving diagnostic features. Data desensitization methods, like differential privacy, should be utilized to protect individual privacy while allowing for collective data patterns to be learned. Additionally, strong encryption standards must be implemented to safeguard stored and transmitted data, ensuring that only authorized users can access sensitive information. Considering decentralized data storage solutions may also help reduce single points of failure and potential attack vectors \cite{roberts2016standardization, ribaric2017identification}.
\section{Conclusion}
In remote regions and developing nations, limited access to high-quality medical resources poses challenges for the precise diagnosis and rehabilitation needs of Parkinson's disease patients. We introduce HypomimiaCoach for hypomimia diagnosis and rehabilitation. Diagnosis use a GNN model to extract AU features to detect hypomimia, provides possibility for digital diagnosis. Rehabilitation helps patients to training their facial muscle scientifically and systematically, with an AU recognition model providing feedback. A formative study of medical knowledge and physicians' clinical experience highlights some of the key findings in the treatment of patients with hypomimia and suggests design goals. In the user study, 7 patients and 10 physicians and 3 rehabilitation therapists participated in the experiment and the digital therapy system received positive feedback in terms of usability, engagement and self-efficacy. By providing targeted feedback and guidance, the system enhances patients' ability to conduct rehabilitation training on their own, which helps to improve their self-management ability and quality of life.
\section{Limitations and Future Work}
There are several limitations to our study that need to be considered. First, this work was done in a hospital in China with a relatively small sample of participants (7). The small sample size is a limitation in itself, as it makes it difficult to generalize our findings to a broader group of Parkinson's disease patients. Details of the system's generalizability to a wider population over a longer period of time are limited. Future work should attempt to involve a wider range of participants with different disease states, but should be mindful of safety and recruitment challenges. In addition, long-term experiments to evaluate the system in rehabilitation settings (e.g., nursing homes) are needed to assess its effectiveness. We can evaluate the system's ability to improve the hypomimia face through a periodic experimental design and data analysis of the mean and variance of the au in conjunction with the mds-updrs scale. The assessment of voice and emotion also requires specialised scales for long-term follow-up ratings. Future research should further explore how to adjust and optimize rehabilitation programs to be more personalized and effective based on patient-specific conditions and feedback. In addition, a dataset could be created to build facial data to make the model more accurate and efficient for future facial interaction studies on a population of Parkinson's hypomimia patients.

\bibliographystyle{ACM-Reference-Format}
\bibliography{reference}
\clearpage

\appendix

\begin{table*}[htbp]
\centering
\caption{Design philosophy based on Chinese-opera}
\label{tab:results}
\begin{tabularx}{\textwidth}{XX}
\toprule
Opera Filter & Conceptual Framework for Filter Design \\
\midrule
'Rolling Lantern'  & The opera 'Rolling Lantern' is a revered masterpiece in Sichuan opera, celebrated for its comedic brilliance, including face-changing and fire-spitting acts. Opera makeup filters are inspired by its humorous style, creating an ambiance that brings laughter and joy to users. \\
\midrule
'Celestial Beauty Scattering Flowers' & The opera 'Celestial Beauty Scattering Flowers' depicts a celestial maiden gracefully carrying a basket of flowers, bestowing blossoms upon the mortal realm, thereby averting disasters. The opera's rhythm transitions from slow to fast, portraying the maiden's joyous emotions as she embarks on her celestial journey. This design aims to imbue users with a sense of delight and happiness. \\
\midrule
'The Case of Chen Shimei' & The opera 'The Case of Chen Shimei' character Bao Zheng embodies justice, unwavering integrity, and resolute defiance of authority. His words resonate with power, inspiring a joyful audience. \\
\bottomrule
\end{tabularx}
\end{table*}

\begin{figure*}[htbp]
    \centering
    \includegraphics[width=0.99\textwidth]{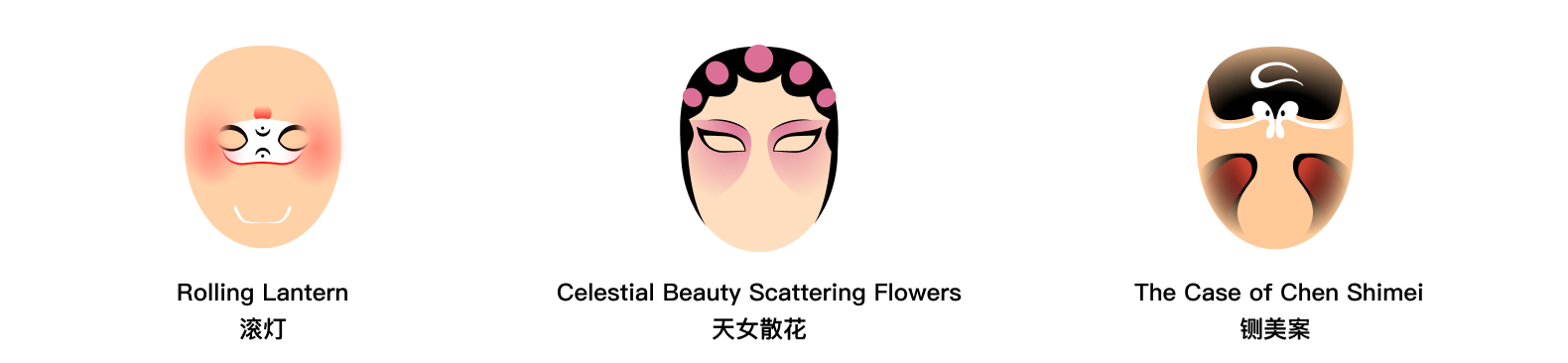} 
    \caption{Design of Chinese opera filters}
    \label{fig11}
\end{figure*}


\begin{figure*}[htbp]
    \centering
    \includegraphics[width=0.99\textwidth]{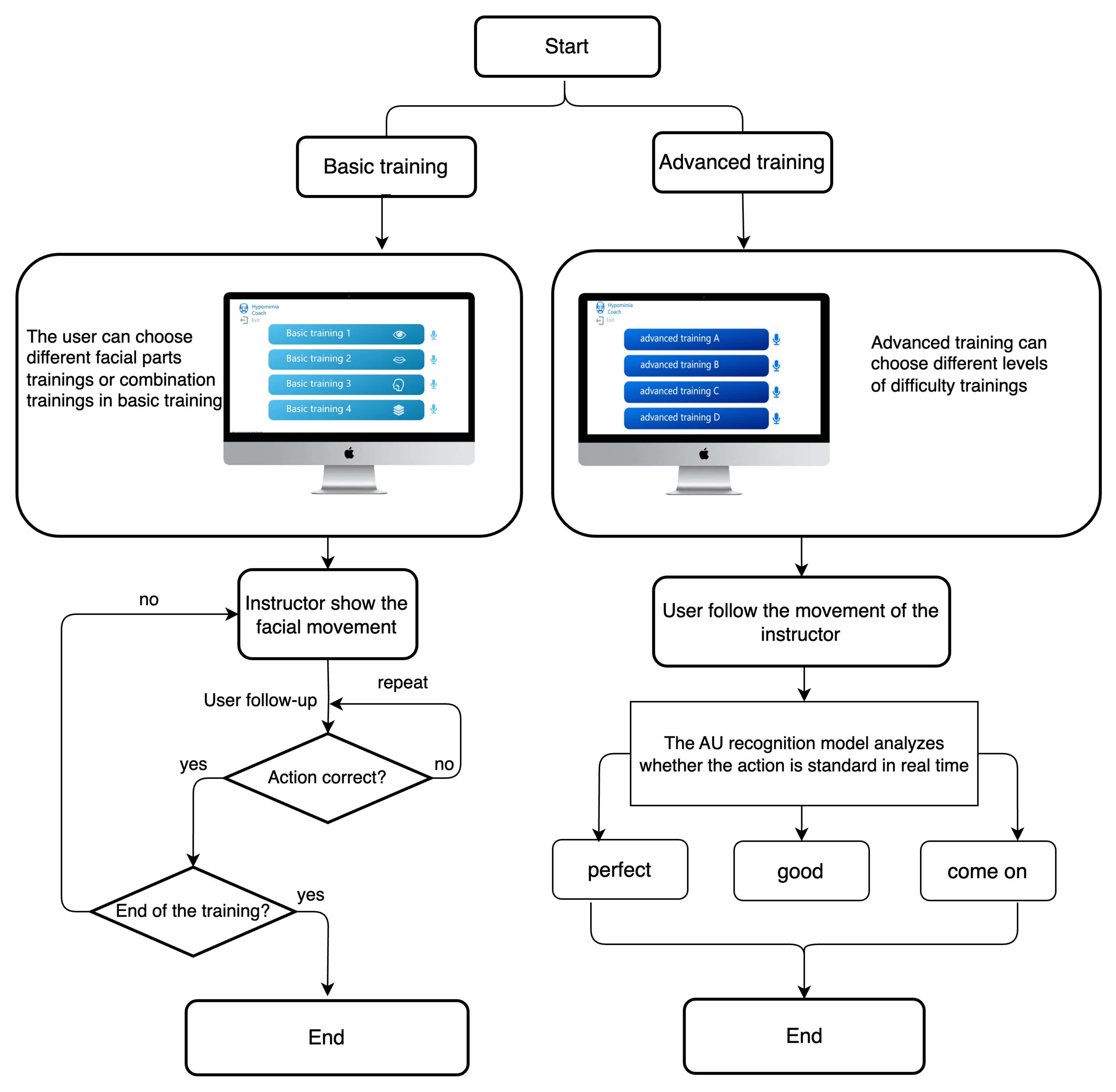} 
    \caption{Procedure of basic training and advanced training. Basic training focuses on specific facial exercises with real-time AI feedback, while advanced training involves operatic exercises tailored to appropriate duration and difficulty.}
    \label{fig:mode}
\end{figure*}

\end{document}